\documentstyle[12pt,epsf]{article}
\def\-{\! - \!}
\begin{document}

\title{Evaluation of the optical conductivity tensor in terms of contour
integrations}
\author{L\'{a}szl\'{o} Szunyogh$^{a,b}$ and Peter Weinberger$^{b}$ \\
\\
$^a$Department of Theoretical Physics, Technical University of Budapest,\\
Budafoki \'ut 8, 1521 Budapest, Hungary \\
$^b$Center for Computational Materials Science, Vienna, Austria \\
Getreidemarkt 9/158, 1060 Wien, Austria}
\date{{\em J. Phys.: Condens. Matter} {\bf 11} (1999) 10451-10458}
\maketitle

\begin{abstract}
For the case of finite life-time broadening the standard Kubo-formula 
for the optical conductivity tensor is rederived
in terms of Green's functions by using contour integrations,   
whereby finite temperatures are accounted for by using
the Fermi-Dirac distribution function. For zero life-time
broadening, the present formalism is related to expressions 
well-known in the literature.
Numerical aspects of how to calculate the corresponding contour integrals 
are also outlined.
\end{abstract}

\section{Introduction}

A general theory of linear response functions has been formulated several
decades ago \cite{kubo}, \cite{wang}, \cite{callaway} that became the basis
of numerous studies of correlation functions, susceptibilities and transport
properties in solid matter. Although early attempts were already devoted 
within this theoretical framework to magneto-optical phenomena, i.e., 
the Faraday and the Kerr effect, \cite{bennett},\cite{wang} it was the rapid
development of both, band structure methods and computational
facilities that brought about the renaissance in this field from the end 
of the 1980's (see \cite
{oppeneer}, \cite{hubert} and refs. therein). This progress was built on the
assumption that the original expression for linear response functions,
derived on the basis of a non-relativistic approach, is still valid when
applying relativistic band structure methods. Recently, this has indeed
been rigorously confirmed by Huhne and Ebert \cite{huhne}.

Within an independent particle picture the interband part of the optical
conductivity tensor at finite temperatures, $\Sigma_{\mu \nu }(\omega )$ is
given by \cite{kubo}, \cite{callaway}

\begin{equation}
\Sigma_{\mu \nu }(\omega )=\frac{\hbar}{i V}\sum_{m,n}
\frac{f(\epsilon _{n})(1\-f(\epsilon_{m}))}{\epsilon _{n}\-\epsilon_{m}}
\left\{ \frac{J_{nm}^{\mu }J_{mn}^{\nu }}
             {\hbar \omega +\epsilon _{n}-\epsilon _{m}+i\delta }
       +\frac{J_{nm}^{\nu }J_{mn}^{\mu }}
             {\hbar \omega +\epsilon _{m}-\epsilon _{n}+i\delta }\right\} 
\; ,
\label{sigma1}
\end{equation}
where $\omega $ is the frequency, $f(\epsilon )$ is the Fermi function, $V$
is the volume of the system, the $J_{nm}^{\mu }=\langle n\!\mid \!J^{\mu
}\!\mid \!m\rangle $ are the matrix elements of the current density operator 
$J^{\mu }$ ($\mu =x,y,z$) with respect to the eigenfunctions $\mid
\!n\rangle $ of the one-electron Hamiltonian corresponding to energy
eigenvalues $\epsilon _{n}$. Life-time broadening effects due to different
scattering processes are partially accounted for by a finite value of $%
\delta =\hbar /\tau \sim 0.03$ Ry, where $\tau $ can be associated with a
relaxation time \cite{oppeneer,hubert}. Such a life-time effect can 
naturally be incorporated into linear response theory either by assuming
an adiabatic switch on of the external field $\phi({\bf r},t)=
\phi({\bf r}) e^{t/\tau}$ \cite{luttinger} or, equivalently, by
convoluting the corresponding frequency dependent conductivity 
for $\delta \rightarrow +0$ by a Lorentzian of a finite halfwidth $\delta$.

Expression (\ref{sigma1}) is usually directly evaluated by applying standard
band structure methods for ordered bulk (three-dimensional translational
invariant) systems by associating the states $\mid n\rangle $ with
(three-dimensional) Bloch-functions \cite{wang},\cite{oppeneer}. This kind
of approach, however, cannot be extended easily to layered systems or
systems with a surface, i.e., to systems with (at best) two-dimensional
translational symmetry. Furthermore, it is totally unsuited to deal with
disorder such as interdiffusion effects in magnetic multilayer systems.

For zero life-time broadening an alternative approach consists of splitting
Eq.~(\ref{sigma1}) into an absorptive and a dispersive part, which in turn
are interrelated via the Kramers-Kronig relations \cite{bennett},\cite
{hubert}. Thus it is sufficient to directly calculate the absorptive part
only, the corresponding Kubo-formula of which can be expressed in terms of
Green's functions as was shown by Butler \cite{butler1} for $\omega =0$, and
was used recently by Banhart \cite{john} for $\omega >0$. 
Even much earlier, Luttinger \cite{luttinger} derived a general formula
of the conductivity tensor which builts on the Green's function 
of the system. However, when
using this method the corresponding energy integrals have to be
performed along the real axis which makes a calculation of $\Sigma _{\mu \nu
}(\omega )$ enormously tedious and also numerically hard to control.

In the present work a suitable extension of Luttinger's formula is given 
in terms of contour integrations that avoids the above mentioned numerical
difficulties, is suitable for applications to magnetic multilayer systems
and for treating alloying effects (interdiffusion etc.) by means of the
Coherent Potential Approximation (CPA).

\section{Contour integrations}

Following Luttinger \cite{luttinger} Eq.~(\ref{sigma1}) can be rewritten as
\begin{equation}
\Sigma_{\mu \nu}(\omega) = \frac{
\sigma_{\mu \nu}(\zeta) \- \sigma_{\mu \nu}(0)}{\zeta} \; ,
\label{Sigma}
\end{equation}
where $\zeta = \omega + i \delta/\hbar$ and
\begin{equation}
\sigma _{\mu \nu }(\zeta)=\frac{i}{V}\sum_{m,n}\frac{f(\epsilon
_{n})-f(\epsilon _{m})}{\hbar \zeta +\epsilon _{n}-\epsilon _{m}}%
J_{nm}^{\mu }J_{mn}^{\nu }\;.  
\label{sigma2}
\end{equation}
For simplicity, as what follows, we denote $\sigma_{\mu \nu}(\zeta)$
by $\sigma_{\mu \nu}(\omega)$ \cite{mahan} the calculation of which we will
concentrate on keeping in mind that we have a finite imaginary
part of the denominator in Eq.~(\ref{sigma2}).
Consider a pair of eigenvalues $\epsilon _{n}$ and $\epsilon _{m}$. For a
suitable contour ${\cal {C}}$ in the complex energy plane (see Fig.~1) the
residue theorem implies 
\begin{eqnarray}
\oint_{{\cal {C}}}\mbox{d}z\frac{f(z)}{(z-\epsilon _{n})(z-\epsilon
_{m}+\hbar \omega +i\delta )} &=&-2\pi i\frac{f(\epsilon _{n})}{\epsilon
_{n}-\epsilon _{m}+\hbar \omega +i\delta }\;+  \label{contour1} \\
&&+\;2i\delta _{T}\sum_{k=-N_{2}+1}^{N_{1}}\frac{1}{(z_{k}-\epsilon
_{n})(z_{k}-\epsilon _{m}+\hbar \omega +i\delta )}\quad ,  \nonumber
\end{eqnarray}
where the $z_{k}=\epsilon _{F}+i(2k-1)\delta _{T}$ ($\epsilon _{F}$ is the Fermi
energy, $k_{B}$ the Boltzmann constant, $T$ the temperature and $\delta
_{T}=\pi k_{B}T$) are the so-called Matsubara-poles. In Eq.~(\ref{contour1}) it
was supposed that $N_{1}$ and $N_{2}$ Matsubara-poles in the upper and lower
semi-plane lie within the contour ${\cal C}$, respectively, i.e., 
\begin{equation}
(2N_{1}-1)\delta _{T}<\delta _{1}<(2N_{1}+1)\delta _{T}\;,  \label{delta1}
\end{equation}
\begin{equation}
(2N_{2}-1)\delta _{T}<\delta _{2}<(2N_{2}+1)\delta _{T}\;,  \label{delta2}
\end{equation}
Eq.~(\ref{contour1}) can be rearranged as 
\begin{equation}
i\frac{f(\epsilon _{n})}{\hbar \omega +\epsilon _{n}-\epsilon _{m}+i\delta }%
=-\frac{1}{2\pi }\oint_{{\cal {C}}}\mbox{d}z\frac{f(z)}{(z-\epsilon
_{n})(z-\epsilon _{m}+\hbar \omega +i\delta )}\;+  \label{contr-nm}
\end{equation}
\[
+\;i\frac{\delta _{T}}{\pi }\sum_{k=-N_{2}+1}^{N_{1}}\frac{1}{%
(z_{k}-\epsilon _{n})(z_{k}-\epsilon _{m}+\hbar \omega +i\delta )}\quad .
\]
Similarly, by choosing a contour ${\cal {C}^{\prime }}$ as shown in Fig.~2
the following expression, 
\begin{equation}
-i\frac{f(\epsilon _{m})}{\hbar \omega +\epsilon _{n}-\epsilon _{m}+i\delta }%
=\frac{1}{2\pi }\oint_{{\cal {C}^{\prime }}}\mbox{d}z\frac{f(z)}{(z-\epsilon
_{m})(z-\epsilon _{n}-\hbar \omega -i\delta )}\;+  \label{contr-mn}
\end{equation}
\[
+\;i\frac{\delta _{T}}{\pi }\sum_{k=-N_{1}+1}^{N_{2}}\frac{1}{%
(z_{k}-\epsilon _{m})(z_{k}-\epsilon _{n}-\hbar \omega -i\delta )}
\]
can be derived. Inserting Eqs.~(\ref{contr-nm}) and (\ref{contr-mn}) into 
Eq.~(\ref{sigma2}) and by closing the contours at $\infty$ and $-\infty$,
$\sigma _{\mu \nu }(\omega )$ is given by 
\begin{equation}
\sigma _{\mu \nu }(\omega )=-\frac{1}{2\pi V}\left\{ \oint_{{\cal {C}}%
}\mbox{d}zf(z)\sum_{m,n}\frac{J_{nm}^{\mu }J_{mn}^{\nu }}{(z-\epsilon
_{n})(z-\epsilon _{m}+\hbar \omega +i\delta )}\;-\right.   \label{sigma-cont}
\end{equation}
\[
\left. \qquad \qquad \qquad \qquad \oint_{{\cal {C}^{\prime }}}\mbox{d}%
zf(z)\sum_{m,n}\frac{J_{nm}^{\mu }J_{mn}^{\nu }}{(z-\epsilon
_{m})(z-\epsilon _{n}-\hbar \omega -i\delta )}\right\} 
\]
\[
\qquad \qquad \;+\;i\frac{\delta _{T}}{\pi V}\left\{
\sum_{k=-N_{2}+1}^{N_{1}}\;\sum_{m,n}\frac{J_{nm}^{\mu }J_{mn}^{\nu }}{%
(z_{k}-\epsilon _{n})(z_{k}-\epsilon _{m}+\hbar \omega +i\delta )}\;+\right. 
\]
\[
\qquad \qquad \qquad \qquad \qquad \left.
\sum_{k=-N_{1}+1}^{N_{2}}\;\sum_{m,n}\frac{J_{nm}^{\mu }J_{mn}^{\nu }}{%
(z_{k}-\epsilon _{m})(z_{k}-\epsilon _{n}-\hbar \omega -i\delta )}\right\}
\quad .
\]
It is now straightforward to rewrite Eq.~(\ref{sigma-cont}) in terms of the
resolvent \cite{pw-book}, 
\begin{equation}
G(z)=\sum_{n}\frac{\mid n\rangle \langle n\mid }{z-\epsilon _{n}}\;,
\label{resolvent}
\end{equation}
such that 
\begin{equation}
\sigma _{\mu \nu }(\omega )=-\frac{1}{2\pi V}\left\{ \oint_{{\cal {C}}%
}\mbox{d}zf(z)Tr\left[ J^{\mu }G(z+\hbar \omega +i\delta )J^{\nu
}G(z)\right] \;-\right. 
\end{equation}
\[
\left. \qquad \qquad \qquad \quad \oint_{{\cal {C}^{\prime }}}\mbox{d}%
zf(z)Tr\left[ J^{\mu }G(z)J^{\nu }G(z-\hbar \omega -i\delta )\right]
\right\} 
\]
\[
\qquad \qquad \;+\;i\frac{\delta _{T}}{\pi V}\left\{
\sum_{k=-N_{2}+1}^{N_{1}}Tr\left[ J^{\mu }G(z_{k}+\hbar \omega +i\delta
)J^{\nu }G(z_{k})\right] \;+\right. 
\]
\[
\qquad \qquad \qquad \qquad \;\;\left. \sum_{k=-N_{1}+1}^{N_{2}}Tr\left[
J^{\mu }G(z_{k})J^{\nu }G(z_{k}-\hbar \omega -i\delta )\right] \right\} \;,
\]
where $Tr$ denotes the trace of an operator. By using the below quantity,
originally introduced by Butler \cite{butler1}, 
\begin{equation}
\widetilde{\sigma }_{\mu \nu }(z_{1},z_{2})=-\frac{1}{2\pi V}Tr\left[
J^{\mu }G(z_{1})J^{\nu }G(z_{2})\right] \;,  \label{sigma-tilde}
\end{equation}
for which the following symmetry relations apply, 
\begin{equation}
\widetilde{\sigma }_{\nu \mu }(z_{2},z_{1})=\widetilde{\sigma }_{\mu \nu
}(z_{1},z_{2}),\quad \widetilde{\sigma }_{\mu \nu }(z_{1}^{*},z_{2}^{*})=%
\widetilde{\sigma }_{\nu \mu }(z_{1},z_{2})^{*}=\widetilde{\sigma }_{\mu \nu
}(z_{2},z_{1})^{*},  \label{sigma-tilde-symm}
\end{equation}
$\sigma _{\mu \nu }(\omega )$ can be written as 
\begin{equation}
\sigma _{\mu \nu }(\omega )= \oint_{{\cal {C}}}%
\mbox{d}zf(z)\widetilde{\sigma }_{\mu \nu }(z+\hbar \omega +i\delta
,z)-\oint_{{\cal {C}^{\prime }}}\mbox{d}zf(z)\widetilde{\sigma }_{\mu \nu
}(z,z-\hbar \omega -i\delta )   \label{sigma-cont1}
\end{equation}
\[
\qquad \qquad -2 i \delta_{T} \left\{
\sum_{k=-N_{2}+1}^{N_{1}}\widetilde{\sigma }_{\mu \nu }(z_{k}+\hbar \omega
+i\delta ,z_{k})+\sum_{k=-N_{1}+1}^{N_{2}}\widetilde{\sigma }_{\mu \nu
}(z_{k},z_{k}-\hbar \omega -i\delta )\right\} \;,
\]
which because of the reflection symmetry for the contours ${\cal C}$ and $%
{\cal C}^{\prime }$ (see Figs.~1 and 2) and the relations 
in Eq.~(\ref{sigma-tilde-symm}) can be transformed to 
\begin{equation}
\sigma _{\mu \nu }(\omega )= \oint_{{\cal {C}}}%
\mbox{d}zf(z)\widetilde{\sigma }_{\mu \nu }(z+\hbar \omega +i\delta
,z)-\left( \oint_{{\cal {C}}}\mbox{d}zf(z)\widetilde{\sigma }_{\mu \nu
}(z-\hbar \omega +i\delta ,z)\right) ^{*}   \label{sigma-cont2}
\end{equation}
\[
\qquad \qquad - 2 i \delta_{T} \sum_{k=-N_{2}+1}^{N_{1}}%
\left\{ \widetilde{\sigma }_{\mu \nu }(z_{k}+\hbar \omega +i\delta ,z_{k})+%
\widetilde{\sigma }_{\mu \nu }(z_{k}-\hbar \omega +i\delta
,z_{k})^{*}\right\} \;.
\]
Eq.~(\ref{sigma-cont2}) displays the central result of the present work: it
compactly expresses the optical conductivity tensor in terms of
contributions of a contour integral and those due to Matsubara poles. The
occurring quantities of the type $\widetilde{\sigma }_{\mu \nu }(z_{1},z_{2})
$ are now exactly of the form such that substitutional disorder (CPA) for
layered systems \cite{butler2,cpa} can be treated. It should be noted that
in the case of site-diagonal terms, see \cite{cpa}, in principle for the
evaluation of the contour integrals also the `irregular' part of the Green's
function has to be included. Since the inhomogeneous CPA for semi-infinite
systems and the corresponding Kubo-Greenwood approach to electric transport
is discussed at length in \cite{cpa}, no further discussion with respect to
layered systems is needed.

\section{Integration along the real axis: the limit of zero life-time 
broadening}

In this Section we give the relationship of Eq.~(\ref{sigma-cont2})
to formulations existing in the literature. For this reason the contour
${\cal C}$ is deformed to the real axis such that
the contributions from the Matsubara poles vanish. By using the 
relations in Eq.~(\ref{sigma-tilde-symm})
Eq.~(\ref{sigma-cont2}) reduces to
\begin{eqnarray}
\sigma _{\mu \nu }(\omega ) &=& \int_{-\infty}^{\infty}
\mbox{d}\epsilon f(\epsilon) \; [
\widetilde{\sigma }_{\mu \nu}
(\epsilon +\hbar \omega + i \delta,\epsilon + i0) -
\widetilde{\sigma }_{\mu \nu}
(\epsilon +\hbar \omega + i \delta, \epsilon \- i0) ] 
\label{sigma-d0-1} \\
&-&\int_{-\infty }^{\infty }\mbox{d}\epsilon f(\epsilon) \: [
\widetilde{\sigma }_{\mu \nu}
(\epsilon \- i0, \epsilon \-\hbar \omega \- i \delta)
\- \widetilde{\sigma }_{\mu \nu}
(\epsilon + i0, \epsilon \-\hbar \omega \- i \delta) ]
  \nonumber 
\end{eqnarray}
\begin{equation}
= - \frac{1}{2 \pi V} \int_{-\infty}^{\infty} 
\mbox{d}\epsilon f(\epsilon ) Tr\left\{
  J^{\mu }G(\epsilon+\hbar \omega + i\delta)
  J^{\nu } \{G^+(\epsilon) \- G^-(\epsilon) \} \right.
\label{sigma-d0-2}
\end{equation}
$$ \left. \qquad \qquad \qquad \qquad \qquad 
+ J^{\mu } \{ G^+(\epsilon) \- G^-(\epsilon) \} 
  J^{\nu } G(\epsilon\-\hbar \omega \- i \delta)
\right\} \; ,
$$
where we introduced up- and down-side limits for the resolvents \cite{pw-book}.
By taking the limit $\delta \rightarrow 0$ Eq.~(\ref{sigma-d0-2})
becomes equivalent to
Eq.~(5.15) of Ref.~\cite{luttinger} for {\bf q}=0,
\begin{equation} 
\sigma _{\mu \nu }(\omega ) = - \frac{1}{2 \pi V} \int_{-\infty}^{\infty}
\mbox{d}\epsilon f(\epsilon ) Tr\left\{
  J^{\mu }G^+(\epsilon+\hbar \omega)
  J^{\nu } \{G^+(\epsilon) \- G^-(\epsilon) \} \right.
\label{sigma-lutt}
\end{equation}
$$ \left. \qquad \qquad \qquad \qquad \qquad \qquad \qquad
+ J^{\mu } \{ G^+(\epsilon) \- G^-(\epsilon) \} 
  J^{\nu } G^-(\epsilon\-\hbar \omega)
\right\} \; .
$$
Note that by shifting in the second term of Eq.~(\ref{sigma-lutt}) 
the argument of integration by $\hbar \omega$ the {\em hermitean} part
of $\sigma _{\mu \nu }(\omega )$ can be expressed as
\begin{equation}
\sigma_{1,\mu \nu}(\omega) \equiv \frac{1}{2} (\sigma_{\mu \nu}(\omega) + 
\sigma_{\nu \mu}(\omega)^*) = - \frac{1}{4 \pi V} 
\int_{-\infty }^{\infty }\mbox{d}\epsilon \;
(f(\epsilon )-f(\epsilon +\hbar \omega )) \times 
\qquad \qquad \qquad \qquad 
\label{sigma1-d0}
\end{equation}
$$ \qquad \qquad \qquad \qquad \qquad \qquad \qquad \qquad Tr\left\{
  J^{\mu } \{ G^+(\epsilon+\hbar \omega) \- G^-(\epsilon +\hbar \omega) \}
  J^{\nu } \{ G^+(\epsilon) \- G^-(\epsilon) \} \right\} \; .
$$
Since quite clearly $\sigma_{1,\mu \nu}(0)=0$, from Eq.~(\ref{Sigma}) we get
$\Sigma_{1,\mu \nu}(\omega) = \sigma_{1,\mu \nu}(\omega)/\omega$ 
as used in Ref.~\cite{john}. 

In order to obtain the correct zero frequency formula one has to take
first the $\omega \rightarrow 0$ limit of Eq.~(\ref{sigma-d0-1})
and then, after inserting into Eq.~(\ref{Sigma}), 
the $\delta \rightarrow 0$ limit has to be performed.
Making use of the analyticity of the Green's functions in the upper and
lower complex semi-planes this leads to \cite{smrcka}
\begin{equation}
\Sigma_{\mu \nu}(0) = 
- \frac{\hbar}{2 \pi V} \int_{-\infty}^{\infty}
\mbox{d}\epsilon \: f(\epsilon) \; Tr\left\{
  J^{\mu } \frac{\partial G^{+}(\epsilon)}{\partial \epsilon}
  J^{\nu} [G^+(\epsilon)\-G^-(\epsilon)] -
  J^{\mu} [G^+(\epsilon)\-G^-(\epsilon)] 
  J^{\mu } \frac{\partial G^{-}(\epsilon)}{\partial \epsilon} 
\right\} \; ,
\end{equation}
which can be integrated by parts.
The corresponding expression at $T = 0$ for the diagonal elements yields 
Butler's original formula \cite{butler1},
\begin{equation}
\Sigma_{\mu \mu}(0)_{(T=0)} = 
- \frac{\hbar}{4 \pi V} Tr\left\{
  J^{\mu }[G^+(\epsilon_F)\-G^-(\epsilon_F)]
  J^{\mu }[G^+(\epsilon_F)\-G^-(\epsilon_F)] \right\} \; .
\end{equation}

\section{Practical evaluation of the contour integrals}

Returning to Eq.~(\ref{sigma-cont2}) we are left with the task to evaluate
the occurring contour integrals in a manner suitable for computational
purposes. This is in particular demanding because, in principle,
integrations from $-\infty $ to $\infty $ are involved. Recalling Fig.~1, it
is essential to note that the integration pathes can be chosen arbitrarily
with the only constraint that the one in the lower complex
semi-plane should lie in the range $0 > Im(z) > -\delta$. 
It is trivial to show that, if the contributions from the core-states 
are not considered, the contour ${\cal C}$
can be closed at $\epsilon _{b}$ (bottom of the valence band) which
lies energetically just below the regime of valence states.
Consequently, the integrals in Eq.~(\ref{sigma-cont2}) can be split up 
as follows 
\begin{equation}
\oint_{{\cal {C}}}\mbox{d}zf(z)\widetilde{\sigma }_{\mu \nu
}(z\pm \hbar \omega +i\delta ,z)=
\label{split1}
\end{equation}
$$
\int_{\epsilon _{b}+i0}^{\infty +i\delta _{1}}\mbox{d}zf(z)%
\widetilde{\sigma }_{\mu \nu }(z\pm \hbar \omega +i\delta ,z)
-\int_{\epsilon_b\-i0}^{\infty \- i \delta_2}\mbox{d}z f(z)
\widetilde{\sigma }_{\mu \nu }(z \pm \hbar \omega +i\delta,z) \;.
$$
Because of the fast decay of the Fermi function $f(z)$ for $Re(z) >
\epsilon_F$ the integrals on the $rhs$ of Eq.~(\ref{split1})
can be terminated at 
$$\epsilon_{u}=\epsilon _{F}+n\,k_{B}T \qquad 
(n\simeq 5-10).$$

Finally, after manipulating the contributions from the
Matsubara poles, see Eq.~(\ref{sigma-cont2}), the optical conductivity
(or rather the zero wave-number current-current correlation function 
\cite{mahan}) can be written as

\begin{equation}
\sigma_{\mu \nu}(\omega) = \sigma^{1}_{\mu \nu}(\omega) + \sigma^{2}_{\mu
\nu}(\omega) + \sigma^{3}_{\mu \nu}(\omega) 
\; ,
\end{equation}
where

\begin{equation}
\sigma _{\mu \nu }^{1}(\omega )=
\int_{\epsilon _{b}+i0}^{\epsilon _{u}+i\delta _{1}}\mbox{d}zf(z)\widetilde{%
\sigma }_{\mu \nu }(z+\hbar \omega +i\delta ,z)-\left( \int_{\epsilon
_{b}+i0}^{\epsilon _{u}+i\delta _{1}}\mbox{d}zf(z)\widetilde{\sigma }_{\mu
\nu }(z-\hbar \omega +i\delta ,z)\right) ^{*} \;,  \label{sigma-1}
\end{equation}

\begin{equation}
\sigma _{\mu \nu }^{2}(\omega )= -
\int_{\epsilon _{b}-i0}^{\epsilon _{u}-i\delta _{2}}\mbox{d}zf(z)\widetilde{%
\sigma }_{\mu \nu }(z+\hbar \omega +i\delta ,z)+\left( \int_{\epsilon
_{b}-i0}^{\epsilon _{u}-i\delta _{2}}\mbox{d}zf(z)\widetilde{\sigma }_{\mu
\nu }(z-\hbar \omega +i\delta ,z)\right) ^{*} \;,  \label{sigma-2}
\end{equation}

and 
\begin{eqnarray}
\sigma _{\mu \nu }^{3}(\omega )=\!\! &-\!\!&\!\! 2 i \delta _{T}
\sum_{k=N_{2}+1}^{N_{1}}\{\widetilde{\sigma }_{\mu \nu }(\epsilon
_{F}+\hbar \omega +i[\delta +(2k-1)\delta _{T}],\epsilon _{F}+i(2k-1)\delta
_{T})  \label{sigma-3} \\
&&\qquad \qquad \left. +\,\widetilde{\sigma }_{\mu \nu }(\epsilon
_{F}-\hbar \omega +i[\delta +(2k-1)\delta _{T}],\epsilon _{F}+i(2k-1)\delta
_{T})^{*}]\right\}   \nonumber \\
&-\!\!&\!\! 2i\delta _{T}\;\;\sum_{k=1}^{N_{2}}\;\left%
\{ \widetilde{\sigma }_{\mu \nu }(\epsilon _{F}+\hbar \omega +i[\delta
+(2k-1)\delta _{T}],\epsilon _{F}+i(2k-1)\delta _{T})\right.   \nonumber \\
&&\qquad \quad \;\; +\widetilde{\sigma }_{\mu \nu }(\epsilon _{F}-\hbar \omega
+i[\delta +(2k-1)\delta _{T}],\epsilon _{F}+i(2k-1)\delta _{T})^{*} 
\nonumber \\[5pt]
&&\qquad \quad \;\; +\widetilde{\sigma }_{\mu \nu }(\epsilon _{F}+\hbar \omega
+i[\delta -(2k-1)\delta _{T}],\epsilon _{F}-i(2k-1)\delta _{T})  
\nonumber \\[5pt]
&&\qquad \quad \;\; +\left. \!\! 
\widetilde{\sigma }_{\mu \nu }(\epsilon _{F}-\hbar
\omega +i[\delta -(2k-1)\delta _{T}],\epsilon _{F}-i(2k-1)\delta
_{T})^{*}\right\} \;.  \nonumber
\end{eqnarray}

Preliminary calculations show that by choosing $\delta \simeq 0.05$ Ryd, $%
\delta _{1}\simeq 0.1-0.15$ Ryd and by using Gaussian-quadrature a total of
30 - 100 energy points (depending on $\hbar \omega $) is sufficient to sample
the above energy integrals. Furthermore, since most of these energy points
lie either below the valence band or far enough away from the real axis (the
energy point closest to the real axis is at $z_{1}=\epsilon _{F}+i\delta _{T}$,
where for $T=300$ K $\delta _{T}\simeq 6$ mRyd), the occurring
Brillouin-zone integrals (see Ref.~\cite{cpa}) can be evaluated by using a
sufficiently small set of ${\bf k}$-points in the surface Brillouin zone,
thus ensuring excellent numerical stability.

\section{Summary}

In the present paper, by using a finite life-time broadening 
(or a finite adiabatic switch-on parameter) 
we reformulated the expression for the (linear
response) optical conductivity tensor 
in terms of contour integrations 
such that Green's functions (resolvents) naturally appear and
finite temperatures are represented by the Fermi-Dirac distribution function. 
For zero life-time broadening the traditional expressions of the
conductivity tensor in terms of the Green's function are recovered. 
Finally, a scheme for calculating the contour integrals that occur, 
in a numerically efficient way, is suggested. 
Although primarily applications to ab-initio calculations of
magneto-optical properties of ordered and disordered thin films in terms of
the fully-relativistic spin-polarized Screened Korringa-Kohn-Rostoker method 
\cite{cpa,szunyogh} are currently under way, it should be noted that the
present method can also be used to calculate other linear response
functions, just as readily as it can easily be extended to the evaluation of
non-linear response functions. 

\newpage
{\large \bf Acknowledgements}
\bigskip

The authors are especially grateful to Prof.
B.L. Gy\"orffy for many stimulating discussions.
This paper resulted from a collaboration partially funded by the Research and
Technological Cooperation between Austria and Hungary
(OMFB-Bundesministerium f\"{u}r Ausw\"{a}rtige Angelegenheiten, Contract
No.  A-35/98).  Financial support was
provided also by the Austrian Science Foundation (Contract No.'s P12146 and
P12352), and the Hungarian National Science Foundation (Contract No.'s OTKA
T024137 and T030240).

\newpage

\begin{figure}
\epsfxsize=12cm
\centerline{\epsfbox{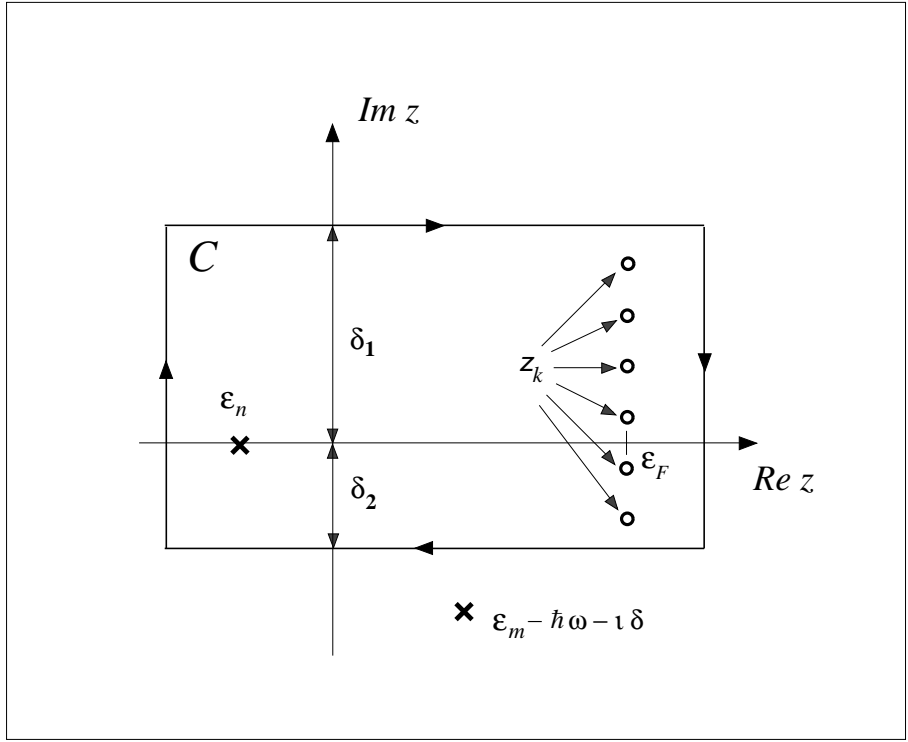}}
\caption{Contour in the complex plane corresponding to the integration
in Eq.~(\protect{\ref{contour1}}).}
\end{figure}

\begin{figure}
\epsfxsize=12cm
\centerline{\epsfbox{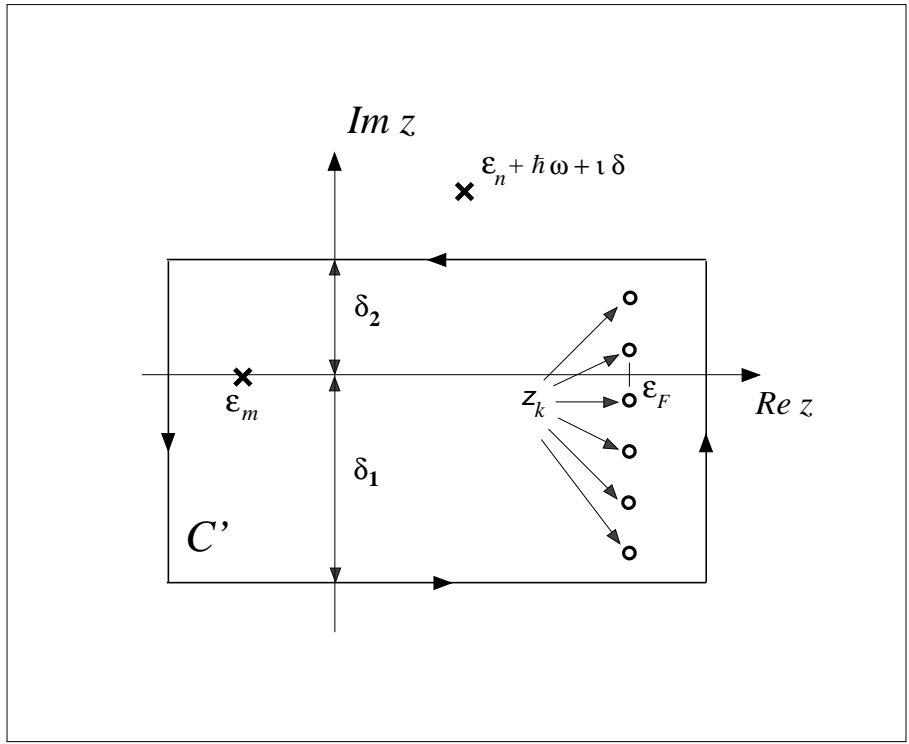}}
\caption{Contour in the complex plane corresponding to the integration
in Eq.~(\protect{\ref{contr-mn}}).}
\end{figure}


\begin{thebibliography}{99}
\bibitem{kubo}  R. Kubo, J. Phys. Soc. Jpn. {\bf 12}, 570 (1957).

\bibitem{wang}  C.S. Wang and J. Callaway, Phys. Rev. B {\bf 9}, 4897 (1974).

\bibitem{callaway}  J. Callaway, {\em Quantum Theory of the Solid State},
part B (New York: Academic, 1974).

\bibitem{bennett}  H.S. Bennett and E.A. Stern, Phys. Rev. {\bf 137}, A448
(1965).

\bibitem{oppeneer}  P. M. Oppeneer and V. N. Antonov, In:
Spin-orbit-influenced Spectroscopies of Magnetic Solids, p. 29 - 47, (Eds.:
H. Ebert and G. Sch\"{u}tz), Springer Verlag 1996.

\bibitem{hubert}  H. Ebert, Rep. Prog. Phys. {\bf 59}, 1665 (1996).

\bibitem{huhne}  T. Huhne and H. Ebert, submitted to Phys. Rev. B (1999).
In there it is claimed that in a fully 
relativistic formalism the so-called diamagnetic contribution to
the conductivity is not explicitely present.

\bibitem{luttinger} J.M. Luttinger, in {\em Mathematical
Methods in Solid State and Superfluid Theory}, Eds. R.C.Clark and
G.H.Derrick (Oliver \& Boyd, Edinburgh, 1967) pp. 157-193.

\bibitem{butler1}  W. H. Butler, Phys. Rev. B {\bf 31}, 3260 (1985).

\bibitem{john}  J. Banhart, Phys. Rev. Lett. {\bf 82}, 2139 (1999).

\bibitem{mahan} In fact $\sigma_{\mu \nu}(\omega) = i \: \Pi_{\mu \nu}({\bf q}
=0,\omega)$, where $\Pi_{\mu \nu}({\bf q},\omega)$ is the {\em current-current}
correlation function. Compare with Eq.~(3.8.9) of G.D. Mahan, {\em
Many-Particle Physics}, Plenum Press, New York, 1990.

\bibitem{pw-book}  P. Weinberger, {\em Electron Scattering for Ordered and
Disordered Matter}, Clarendon Press, 1990.

\bibitem{butler2}  W. H. Butler, X.-G. Zhang and D.M.C. Nicholson, J. Appl.
Phys. {\bf 76}, 6808 (1994); W.H. Butler, X.-G. Zhang, D.M.C. Nicholson and
J.M. MacLaren, Phys. Rev. B {\bf 52}, 13399 (1995).

\bibitem{cpa}  P. Weinberger, P.M. Levy, J. Banhart, L. Szunyogh, and B.
\'{U}jfalussy, J. Phys.: Condens. Matter {\bf 8}, 7679 (1996).

\bibitem{szunyogh}  L. Szunyogh, B. \'{U}jfalussy, and P. Weinberger, Phys.
Rev. B {\bf 51}, 9552 (1995).

\bibitem{smrcka} L. Smr\v{c}ka and P. St\v{r}eda, J. Phys. C {\bf 10}, 
2153 (1977).
\end{thebibliography}
\end{document}